\documentclass[fleqn,twoside]{article}
\usepackage{espcrc2}
\usepackage{amsfonts, amssymb}

\usepackage{graphicx}
\usepackage{rotating}

\newcommand{\AmS}{{\protect\the\textfont2
  A\kern-.1667em\lower.5ex\hbox{M}\kern-.125emS}}

\title{Superstring Field Theory Action Including Massless Fermions}

\author{Luciano Barosi \address[DFMA]{Departamento de F\'\i sica Matem\'atica-IFUSP \\ 
        C.P. 66.318, 05315-970, S\~ao Paulo, SP Brasil }%
        \thanks{Work supported by FAPESP grant 98/00452-7}}

\begin{document}

\begin{abstract}
Using Berkovits' Superstring Filed Theory action including the Ramond
sector, we calculate the contribution to this action coming from the
tachyon and massless fermions from both $GSO$ sectors. Some features of
the action are discussed in the end, settling the ground for a more
systematic treatment of spacetime fermions in superstring field theory.
\vspace{1pc}
\end{abstract}

\maketitle

\section{Introduction}

During the 80's a string field theory was developed for the open bosonic
string theory, following the ideas of witten for the role of
non-commutativity in the string theory product (witten's midpoint
interaction) and the development of BRST techniques in string theory
\cite{Witten1}.

The natural generalization of these ideas to the case of the superstring
theory didn't achieved the success of the bosonic theory due to the
divergences at the tree level for the classical action, a consequence of
picture changing operators appearing explicitly in the action
\cite{Wendt1}. The interest on this kind of theories declined
substantially, mainly because of the lack of any significant progress,
till Sen's conjecture on the role of tachyon in a system of unstable
D-branes \cite{Sen1}, arguing that the tachyon should have a minimum for
its potential at the point where the system undergoes a decay,
establishing a value for the minimum of the tachyon potential as the value
of the tension of the original unstable D-brane system.

String Field Theory is the natural setting for studying the tachyon
condensation and so we have seen a renewal of interest on this theory.

About the same time, Berkovits proposed an action for the NS sector of
superstring field theory which doesn't suffer from the divergences of the
previous formulation \cite{Berkovits2}. This action is based in an
embedding of $N=1$ superconformal algebra into an $N=4$ algebra, and is
explicitly supersymmetric in four dimensions. The Ramond sector was
included last year \cite{Berkovits7} in an action with three string fields that will be
reviewed latter in this talk.

Calculations of the tachyon potential have been done and they show a good
agreement with Sen's conjecture, but this calculations only require the NS
sector \cite{Berkovits2}. Up to now there's no study on
the Ramond sector.

In this work we are going to consider the action of \cite{Berkovits7} for
the superstring field action with manifest $N=1$, $D=4$ supersymmetry. We
are going to include the GSO(-) massless sector and the Tachyon and
calculate the relevant terms of the action, at the end we recombine the
terms in order to write an explicitly $D=10$ Lorentz invariant action. The
pure GSO(+) sector is super-Yang-Mills in 10-dimensions and we are not
going to write it.


The Ramond sector is important in order to understand supersymmetry
related issues regarding tachyon condensation.  If a superstring theory on
an unstable D-brane undergoes tachyon condensation either stable D-branes
(via solitonic solutions of the Tachyon profile) or a true closed string
vacuum (in the annihilation of all the $D-\bar D$ system) should appear,
these theories are supersymmetric and so the original theory should have a
hidden SUSY (in a broken phase), it's just natural to think of GSO(-)
fermions as goldstinos for the broken SUSY.

There are indeed some arguments from Yoneya which support these statements
\cite{Yoneya2}. Also, if the massless GSO(-) fermions are
to be considered as goldstinos, there should be a non-linear supersymmetry
realized in the broken phase, in which the goldstinos transform under the
usual SUSY transformation with a shift due to a Fayet-Iliopoulos term,
which is to be canceled in the tachyonic vacuum, where the linear
supersymmetry is restored.

In the next session we review the hybrid formalism, used to describe the
superstring field theory action in the following session, emphasizing the
issues regarding vertex operators and inclusion of $GSO(-)$ sector. In
section 4 we sketch the computations and discuss some perspectives at the end.

\section{Hybrid Formalism}

\subsection{Hybrid Variables}
This formalism enables a formulation of superstrings in an explicitly D=4
Super-Poincaré invariant manner (for the $GSO(+)$) sector. In order to
understand some of its features, one should consider breaking of the full
10-dimensional Lorentz invariance of the superstrings, by complexifying 6
of the bosonic coordinates (and analogously to the fermionic variables):
\begin{equation}
  \label{eq:complex}
  x^{\pm j} = \frac{1}{\sqrt{2}} (x^{2j+2}\pm x^{2j+3}), \quad j=1,2,3.
\end{equation}

Following the lines of \cite{Berkovits4}, we use the following set of
variables to describe the superstring, including some superfield
coordinates $\theta$ and the conjugate momenta $p$:
\begin{equation}
\left[\;\; X^m,\;\theta^{\alpha},\;\bar\theta^{\dot\alpha},\;
p_{\alpha},\; \bar{p}_{\dot\alpha}, \;\rho,\;\Gamma^{\pm j},\; X^{\pm j}\;\;
\right] \label{3} \nonumber 
\end{equation}
here $X^m$ is the $D=4$ spacetime vector world
sheet boson, $m=0,1,2,3$;\\
$\theta^{\alpha}$ is the $D=4$ spacetime chiral spinor worldsheet
fermion, $\alpha=1,2$; \\
$\bar{\theta}^{\dot\alpha}$ is the space-time anti-chiral spinor
world
sheet fermion, $\dot\alpha=1,2$; \\
$\rho$ is the world sheet chiral boson;\\
$\Gamma^{\pm j}$ are internal world sheet
fermions, $j=1,2,3$; \\
$X^{\pm j}$ are internal world sheet bosons,
$j=1,2,3$.\\

All these fields have free field OPEs:
\begin{eqnarray}
X^m(z)X^n(w)   & = & -\eta^{mn} \ln(z-w),
\nonumber\\
p_{\alpha}(z)\theta^{\beta}(w) & = &
\frac{\delta^{\beta}_{\alpha}}{z-w}, \nonumber
\\
\rho(z)\rho(w) & = & -\ln(z-w), \nonumber\\
\Gamma^{+i}(z) \Gamma^{-j}(w) & = &
\frac{\delta^{ij}}{(z-w)}, \\
X^{+i}(z)X^{-j}(w) &=& -\delta^{ij}\ln(z-w).\label{opes}\nonumber
\end{eqnarray}
and the relation of these variables to the usual RNS variables, written in
the $SO(3,1)$ invariant form:
\begin{equation}
[\; \underbrace{x^{m}, \; \psi^{m},\;b,\; c,\; \xi,\;\eta,\;
\phi}_{\mbox{{\tiny $D=4$ space-time matter + ghosts}}} ,\;\;
\underbrace{ x^{\pm j},\; \psi^{\pm j}}_{internal} \;], \label{2}
\nonumber
\end{equation}
is given by the following set of relations:
\begin{eqnarray}
X^m &=& e^{R+\frac{1}{2}U} (x^m)e^{-R-\frac{1}{2}U},
\nonumber
\\
\theta^{\alpha} & = & e^{\frac{\phi}{2}}\Sigma^{\alpha}
e^{-\frac{H}{2}},  \nonumber  \\
\bar{\theta}^{\dot{\alpha}} & = & c\xi
e^{-\frac{3}{2}\phi}\Sigma^{\dot\alpha} e^{\frac{H}{2}},
     \nonumber\\
p_{\alpha} &=& e^{R+\frac{U}{2}}
(e^{-\frac{\phi}{2}}\Sigma_{\alpha} e^{\frac{H}{2}})
e^{-R-\frac{1}{2}U},
       \nonumber\\
\bar{p}_{\dot\alpha} &=& e^{R+\frac{1}{2}U} (b\eta
e^{\frac{3\phi}{2}}\Sigma_{\dot\alpha} e^{-\frac{H}{2}})
e^{-R-\frac{1}{2}U},
      \nonumber\\
\partial \rho & = & 3\partial\phi - cb -
2\xi\eta -\partial H,
      \nonumber\\
X^{+j} &=& e^{R+\frac{1}{2}U} ( x^{+j})e^{-R-\frac{1}{2}U},
       \nonumber\\
X^{-j} &=& e^{R+\frac{1}{2}U} ( x^{-j})e^{-R-\frac{1}{2}U},
      \nonumber\\
\Gamma^{+j} &=& e^{R+\frac{1}{2}U} ( \xi
e^{-\phi}\psi^{+j})e^{-R-\frac{1}{2}U},
      \nonumber\\
\Gamma^{-j} &=& e^{R+\frac{1}{2}U} ( \eta
e^{\phi}\psi^{-j})e^{-R-\frac{1}{2}U},\label{map}
\end{eqnarray}
where
\begin{eqnarray}
R &=& \int dz \;c \xi e^{-\phi}i\psi^{-j}\partial
x^{+j},                  \nonumber \\
U&=& \int dz [\; c\xi\; e^{-\phi}i\psi^m\partial x_m \nonumber\\ &&
+\frac{1}{2}(\partial\phi+\partial H)c\partial c \xi\partial\xi
e^{-2\phi}]),\nonumber
\end{eqnarray}
the chiral boson $H$ is defined by $\partial H = :~\psi^{+j}\psi^{-j}:$
and the fields $\Sigma$ are spin fields made up by bosonization of the
4-dimensional world-sheet fermions, as explained in \cite{Berkovits4}.

\subsection{Vertex operators}
The vertex operators may be constructed directly in the hybrid formalism
by using its rich superconformal structure. The $GSO(+)$ sector was
already contained in \cite{Berkovits4}, and the $GSO(-)$ vertices for
Tachyon and massless fermions were constructed in detail in
\cite{LB03}. 

The goal of this work is to compute the superstring field theory action
contribution due to these $GSO(-)$ fields, which can be obtained from the
RNS vertices by using the mapping in eqs.(\ref{map}) Since we're going to
obtain vertices with manifest 4-dimensional Lorentz invariance, is
convenient, prior to the exhibition of the vertex operators, to show how
the 10-dimensional fields split in different fields.

In 10-dimensions, the superstring spectrum has a massless vector and a
chiral fermion in the $GSO(+)$ sector and the Tachyon and an anti-chiral
fermion in the $GSO(-)$ sector. Using the Lorentz breaking pattern:
\begin{eqnarray}
\label{lorbreak}
  \nonumber \mathrm{SO(9,1)}&\rightarrow&
\mathrm{SO(3,1)}\times \mathrm{SO(6)},\\
  \mathrm{SO(9,1)}&\rightarrow&
\mathrm{SO(3,1)}\times
  \mathrm{SU(3)}\times \mathrm{U(1)}.
\end{eqnarray}
the 10-dimensional representations split according to:
\begin{eqnarray*}
  10 &\Rightarrow& 4 \quad\oplus\quad 3 \quad\oplus\quad \bar 3 \\
  && A_m \quad\quad \phi^{-i} \quad\quad \bar\phi^{+i}\\
  16 &\rightarrow& (2,1_{+3}) \oplus (2,3_{-1}) \oplus (2', 1_{-3}) \oplus (2',
  \bar 3_{+1})\\
  && \chi_{(+)\alpha} \quad\quad (\psi_\alpha)^{-i} \quad\quad
  \bar\chi_{(+)}^{\dot\alpha} \quad\qquad (\bar\psi^{\dot\alpha})^{+i} \\
  16' &\rightarrow& (2',1_{+3}) \oplus (2',3_{-1}) \oplus (2, 1_{-3}) \oplus (2,
  \bar 3_{+1})\\
  && \bar\chi_{(-)}^{\dot\alpha} \quad\qquad (\bar\lambda^\alpha)^{-i} \quad\quad
  \chi_{(-)\alpha} \quad\quad (\lambda_{\alpha})^{+i} 
\end{eqnarray*}
where primed quantities mean anti-weyl representations, the subscript is
the $U(1)$ charge and the number in parenthesis are the dimensions of the
representations. 
 The vertex operators, obtained by the mapping are summarized in the table
 \ref{tab:1}\footnote{In this table $\Lambda^{++++}$ is a spin field of hybrid variables, from the
bosonized $\theta$ and $p$, as explained in \cite{LB03}}.
\begin{table}[htbp]
  \centering
\[
\begin{array}{ll}
\hline
RNS & Hybrid  \\
\hline
        c\partial c \xi\partial\xi
         e^{-\frac{5}{2}\phi}\Sigma^{\alpha}e^{H/2}\chi^+_{\alpha}
       & \theta^{\alpha}(\bar{\theta})^2 \chi^+_{\alpha}
       \\
       c\xi
         e^{-\frac{1}{2}\phi}\Sigma^{\dot{\alpha}}e^{-H/2}
         \bar{\chi}^+_{\dot\alpha}
       & \bar{\theta}^{\dot\alpha} (\theta)^2\bar{\chi}^+_{\dot\alpha}
       \\
       c\xi e^{-\phi}\psi^m A_m
       & \theta^{\alpha}\sigma^m_{\alpha\dot\alpha}
         \bar{\theta}^{\dot\alpha} A_m
       \\
        c\xi e^{-\phi}t
       & \Lambda^{++++} t
       \\
       z^{\frac{1}{2}}c\xi
         e^{-\frac{1}{2}\phi}\Sigma^{\alpha}e^{-H/2}\chi^-_{\alpha}
       & \theta^{\alpha} \Lambda\chi^-_{\alpha}
       \\
        z^{\frac{1}{2}}c\partial c \xi \partial\xi
         e^{-\frac{5}{2}\phi}\Sigma^{\dot\alpha}e^{H/2}
         \bar{\chi}^-_{\dot\alpha}
       & \bar{\theta}^{\dot\alpha} \Lambda \bar{\chi}^-_{\dot\alpha}
       \\
\hline
        c\xi e^{-\frac{1}{2}\phi} \Sigma^{\alpha} \Xi^{+j}
         \lambda^+_{+j\alpha}
       & e^{\rho}\Gamma^{+j}\theta^{\alpha}(\bar{\theta})^2
         \lambda^+_{+j\alpha}
       \\
        c\xi e^{-\phi}\psi^{-j}A_{-j}
       & e^{\rho}\Gamma^{+j}(\bar{\theta})^2 A_{-j}
       \\
        z^{\frac{1}{2}}c\xi
         e^{-\frac{1}{2}\phi}\Sigma^{\dot{\alpha}}\Xi^{+j}
         \bar{\lambda}^-_{+j\dot\alpha}
       & e^{\rho} \Gamma^{+j} \bar{\theta}^{\dot\alpha} \Lambda
         \bar{\lambda}^-_{+j\dot\alpha}
       \\
\hline
        c\partial c \xi\partial \xi
         e^{-\frac{5}{2}\phi} \Sigma^{\dot\alpha} \Xi^{-j}
         \bar{\lambda}^+_{-j\dot\alpha}
       & e^{-\rho} \Gamma^{-j} \bar{\theta}^{\dot\alpha} (\theta)^2
         \bar{\lambda}^+_{-j\dot\alpha}
       \\
        c\xi e^{-\phi}\psi^{+j}A_{+j}
       & e^{-\rho}\Gamma^{-j}(\theta)^2A_{-j}
       \\
       z^{\frac{1}{2}}c\partial c \xi\partial\xi
         e^{-\frac{5}{2}\phi}\Sigma^{\alpha}\Xi^{-j}\lambda^-_{-j\alpha}
       & e^{-\rho}\Gamma^{-j} \theta^{\alpha} \Lambda\lambda^-_{-j\alpha}
       \\
\hline
\end{array}
\]
  \caption{Hybrid vertex operators from RNS.}
  \label{tab:1}
\end{table}

\section{SSFT Action}

Using these hybrid variables one can construct an $N=4, c=6$
superconformal algebra with the following generators:
\begin{eqnarray}
\label{n=4} L &=& -\frac{1}{2} \partial X^m\partial X_m
-p_{\alpha}\theta^{\alpha} -
{\bar{p}}_{\dot\alpha}{\bar\theta}^{\dot\alpha}
-\partial\rho\partial\rho \nonumber\\&&-\frac{1}{2}\partial^2\rho
+L_c,\nonumber
\\
G^+&=& \underbrace{e^{\rho}(d)^2}_{G^+_0} +
\underbrace{\Gamma^{-j}\partial X^{+j}}_{G^+_{-\frac{1}{3}}},
\nonumber \\
\tilde{G}^+&=& \underbrace{\epsilon^{jkl} \Gamma^{-j}
\Gamma^{-k}\Gamma^{-l}e^{-2\rho}(\bar d)^2}_{G^+_{-1}} +\nonumber\\&&
\underbrace{e^{-\rho}\epsilon^{jkl} \Gamma^{-j}\Gamma^{-k}\partial
X^{-l}}_{G^+_{-\frac{2}{3}}},
\nonumber\\
G^-&=& \underbrace{e^{-\rho}(\bar d)^2}_{G^-_{0}}
+\underbrace{\Gamma^{+j}\partial X^{-j}}_{G^-_{\frac{1}{3}}},\\
\tilde{G}^-&=& \underbrace{\epsilon^{jkl} \Gamma^{+j}
\Gamma^{+k}\Gamma^{+l}e^{+2\rho}( d)^2}_{G^-_1} ++\nonumber\\&&
\underbrace{e^{\rho}\epsilon^{jkl} \Gamma^{+j}\Gamma^{+k}\partial
X^{+l}}_{G^-_{\frac{2}{3}}},     \nonumber\\
J_{gh} &=& \partial \rho +
\Gamma^{-j}\Gamma^{+j},       \nonumber \\
J^{++} &=& e^{-\rho-\Gamma^{-j}\Gamma^{+j}},
   \nonumber\\
J^{--} &=& e^{\rho+\Gamma^{-j}\Gamma^{+j}},
      \nonumber
\end{eqnarray}
where $ L_c$ is the stress tensor for the internal variables
$\Gamma^{\pm
  j},\;X^{\pm j}$, the underbraces show the terms with definite C-charge
$(C=\int \Gamma^{+j}\Gamma^{-j})$ which helps constructing the action, and
\begin{eqnarray}
d_{\alpha}(z)&=&p_{\alpha}(z)+
\frac{i}{2}\bar{\theta}^{\dot\alpha}
\sigma^m_{\alpha\dot\alpha}\partial X_m(z) \\&&-\frac{1}{4}(\bar
\theta)^2\partial\theta_{\alpha}(z)+
+\frac{1}{8}\theta_{\alpha}\partial(\bar\theta)^2(z) \label{d}
\end{eqnarray}
is the current associated to the $N=1$, $D=4$ supersymmetric
covariant derivative $D_{\alpha}$. 

Berkovits superstring field theory action, developed in \cite{Berkovits7}
is:
\begin{eqnarray*}
\nonumber    &S = {\langle (e^{-\Phi} G_{-1} e^\Phi)(e^{-\Phi} G_0 e^\Phi)}\\&+ {(e^{-\Phi}
  G_{\frac{2}{3}} e^\Phi)(e^{-\Phi} G_{-\frac{1}{3}} e^\Phi) + }\\ \nonumber
\nonumber  & {\int_0^1 dt (e^{-\hat\Phi}\partial_t e^{\hat \Phi}) \left( \{e^{-\hat\Phi}G_{-1}
  e^{\hat \Phi}, e^{-\hat\Phi}G_0 e^{\hat \Phi} \}\right) }\\\nonumber
& {\int_0^1 dt (e^{-\hat\Phi}\partial_t e^{\hat \Phi}) \left(\{e^{-\hat\Phi} G_{-\frac{2}{3}}
  e^{\hat \Phi}, e^{-\hat\Phi}G_{-\frac{1}{3}} e^{\hat \Phi} \}\right)  }\\
\nonumber  & {-e^{- \Phi} {\bar\Omega} e^{ \Phi}  \Omega +  {\bar\Omega} e^{ \Phi}
  G_{-\frac{2}{3}}e^{- \Phi}  +  \Omega e^{- \Phi}
  G_{-\frac{1}{3}}e^{+ \Phi} \rangle_D} \\\nonumber
  & {- \langle \frac{1}{2}  {\bar\Omega}
    G_{-\frac{1}{3}}  {\bar\Omega}+ \frac{1}{3}{\bar\Omega}^3
  \rangle_{\bar F}}\\\nonumber
& +{\langle \frac{1}{2}  {\Omega}
    G_{-\frac{2}{3}}  {\Omega} +  \frac{1}{3}{\Omega}^3 \rangle_{F}}
\end{eqnarray*}

where the operators $G$ appearing in this action are defined in
eq.(\ref{n=4}). We should also note that there are three different
correlators. The D-correlator is defined in the large Hilbert
Space of the superstring. The $F$ and $\bar F$ correlators are
chiral (anti-chiral) subspaces defined with the trivial cohomology
pieces of the G operator, $G_{-1}$ and $G_0$.

The string fields appearing in the action are defined in such a way that
$\Phi$ has zero C-charge, $\Omega\equiv G_{-1}\Psi$, is a chiral field
obtained from the string field $\Psi$ with C-charge $1/3$ and $\bar \Omega
\equiv G_0 \bar \Psi$ is an anti-chiral string field obtained from a
string field $\bar \Psi$ with C-charge $-1/3$. We are using the notation
$\tilde\Phi=t\Phi,\;\;0\leq t\leq 1$ for the WZW part. Note that with this
notation all the 4-dimensional fields are in $\Phi$, while the fields in
the other string fields are dependent of the ``internal'' part.

All products between string fields are the midpoint interaction, calculated
according to the following prescription:
\begin{eqnarray}
  \langle V_1 V_2 \cdots V_N \rangle
&=& \left( -\frac{4i}{N}\right)^{\sum_{k=1}^N h_k} e^{\frac{2\pi
  i}{N}\sum_{l=1}^N h_l(l-1)} \nonumber\\&&\left\langle \Pi_{j=1}^N V_j\left(
e^{\frac{2\pi i}{N}(j-1)} \right)\right\rangle
\end{eqnarray}
where $V_i$ here are conformal primary fields with conformal
weight $h_i$.

The non-vanishing norms for the three subspaces are obtained by background
charge cancellation \cite{Berkovits7}:
\begin{eqnarray*}
  \langle \frac{1}{24} \theta^2 {\bar\theta}^2 e^{-\rho} \epsilon^{ijk} \Gamma^{-i}\Gamma^{-j}\Gamma^{-k}
  \rangle_D&=&1 \\
  \langle \frac{1}{24} {\bar\theta}^2 \epsilon^{ijk} \Gamma^{-i}\Gamma^{-j}\Gamma^{-k}\rangle_{\bar F} &=& 1 \\
 \langle \frac{1}{4}e^{-3\rho} \theta^2 (\Gamma \cdot\partial\Gamma)^3 \rangle_F &=&1  
\end{eqnarray*}
There's a small subtlety when including $GSO(-)$ string fields in the
action, since these fields, by their different commutativity properties,
may spoil ciclicity of the correlators. The way out of this is considering
a suitable tensor product of the string fields with $2\times 2$ matrices
(Pauli Matrices), \cite{Ohmori1}. 
\begin{eqnarray*}
  \Phi &=& \Phi^+ \otimes I + \Phi^- \otimes \sigma_1\\
  G &=& G\otimes \sigma_3
\end{eqnarray*}
and analogous relations for the other string fields and operators. 

The vertex operators discussed in the previous section may be organized in
superfields (for the $GSO(+)$ sector), since the hybrid formalism is D=4
super-poincare invariant, Doing so we have the following set of string
fields for the tachyon and massless fermions:
\begin{eqnarray}
  \Phi^+ &=& v(x,\theta,\bar\theta) \\ 
  \Psi^+ &=& e^\rho \Gamma^{+j} \omega^{-j}(x,\theta) \bar\theta^2 \\
  \bar\Psi^+ &=& e^{-\rho} \Gamma^{-j} \bar\omega^{+j}(x,\bar\theta) \theta^2 \\ 
  \Phi^- &=& \!\!\!\!\Lambda^{++++}T(x) + \Theta^\alpha \chi_\alpha(x)
  +\bar\Theta^{\dot\alpha}\bar\chi_{\dot\alpha}(x)\\  
  \Psi^- &=& e^\rho \Gamma^{+j} \bar\Theta^{\dot\alpha} \bar\lambda_{\dot\alpha}^{-j}(x) \\ 
  \bar\Psi^- &=& e^{-\rho} \Gamma^{-j} \Lambda^{\alpha} \lambda_{\alpha}^{+j}(x) 
\end{eqnarray}
where $v(x,\theta,\bar\theta)$ is an $N=1, D=4$ vector superfield,
$\omega(x,\theta)$ is a chiral superfield and $\Theta \equiv \theta
\Lambda^{++++}$. Similar relations hold for the barred quantities.

Expanding the non-polynomial string field action up to terms with four
string fields \footnote{since for the vertices we are considering all
  other terms vanish due to background charge cancellation} and plugging
the vertices in this action, it's possible to compute all the correlators
in conformal field theory, getting an action with manifest 4-dimensional
Lorentz invariance. From the knowledge of the Lorentz breaking pattern in
eq(\ref{lorbreak}), it's still possible to recover a 10-dimensional
Lorentz invariant action for these fields. The details of the computation,
although a bit straightforward, may be found in \cite{LB03}.

The classical action for the Tachyon and massless fermions in both GSO
sectors is:
\begin{eqnarray}
  \nonumber S&=& \mathrm{Tr} \int d^{10} x
\left[ \frac{1}{4} F^{\mu\nu}F_{\mu\nu}+ T\Box T +\frac{1}{2}T^2\right.\\&&
  +[T,A^{\mu}][T,A_{\mu}]-T^4
 \nonumber \\
&& + \chi^{a\;+} \gamma^{\mu}_{ab} \left(\partial_\mu
\chi^{b\;+}+[A_{\mu},\chi^{b\;+}] \right) + \nonumber\\&&\chi^{a\;-}
\gamma^{\mu}_{ab} \left(\partial_\mu
\chi^{b\;-}+[A_{\mu},\chi^{b\;-}] \right)
\nonumber\\
&&\left. + 3^\frac{3}{4} T (\chi^{a\;+} \chi^-_{a})+\cdots \right]
\end{eqnarray}
where $F_{mn}$ is the field-strength for the gauge-field $A_m$,
$T$ is the tachyon field, $\chi^{+}_{a}$, $\chi^-_{a}$ are the
massless fermions coming from the $GSO(+)$ and $GSO(-)$ sector
that are 10-dimensional chiral and anti-chiral respectively. 

We should comment on some features of this action. First, the $GSO(+)$
part was not calculated in the work; this action have higher derivative
terms coming from the OPEs of the spacetime fields inserted in different
points in the correlators and there exist a trilinear coupling among
fermions from both $GSO$ sectors and the Tachyon, signalling for the
possibility of the massless $GSO(-)$ fermions to be goldstinos.

\section{Conclusions}

We performed a first non-trivial calculation with the Berkovits'
superstring field theory action, which includes the Ramond sector, and
it's rather clear that the hybrid formalism, together with this approach
to superstring field theory may be very fruitful in the future. The study
of supersymmetry breaking related issues regarding tachyon condensation is
poorly understood and rely mostly upon guesswork and intuitive
arguments. We hope the setup presented here may help the way into a more
systematic treatment of this subject.

In particular, it's possible to construct a Boundary String field Theory \cite{BSFT}
in the hybrid formalism, which is currently under development. The
insights from both approaches may be as fruitful as it proved to be in the
study of the tachyon potential \cite{Zieb}.

This formalism is also particularly well suited to discuss Ramond-Ramond
backgrounds, which are still far from being well understood.

\section{Acknowledgements}

LB would like to thank Carlos Tello, with whom part of this work was
developed; Nathan Jacob Berkovits and Brenno Vallilo for inumerous
discussions.

\end{document}